\documentclass[prl, twocolumn, superscriptaddress, longbibliography]{revtex4-1}
\usepackage{amsmath}
\usepackage{amssymb}
\usepackage{graphicx}
\usepackage[caption=false, position=top, singlelinecheck=off, justification=raggedright]{subfig}
\usepackage{color}
\usepackage{pst-node}
\usepackage[unicode]{hyperref}
\hypersetup{unicode=true, colorlinks=true, linkcolor=blue, citecolor=blue}

\begin{document} 

\title{Observation of Brownian Motion of a Bose-Einstein Condensate}
\author{Xiao-Qiong Wang}
\affiliation{Shenzhen Institute for Quantum Science and Engineering and Department of Physics, State key laboratory of quantum functional materials, and Guangdong Basic Research Center of Excellence for Quantum Science, Southern University of Science and Technology, Shenzhen 518055, China}
\author{Rui-Lang Zeng}
\affiliation{Shenzhen Institute for Quantum Science and Engineering and Department of Physics, State key laboratory of quantum functional materials, and Guangdong Basic Research Center of Excellence for Quantum Science, Southern University of Science and Technology, Shenzhen 518055, China}
\author{Zi-Yao Zhang}
\affiliation{Shenzhen Institute for Quantum Science and Engineering and Department of Physics, State key laboratory of quantum functional materials, and Guangdong Basic Research Center of Excellence for Quantum Science, Southern University of Science and Technology, Shenzhen 518055, China}
\author{Chushun~Tian}
\affiliation{CAS Key Laboratory of Theoretical Physics and Institute of Theoretical Physics,
Chinese Academy of Sciences, Beijing 100190, China}
\author{Shizhong Zhang}
\affiliation{Department of Physics and Hong Kong Institute of Quantum Science and Technology, The University of Hong Kong, Hong Kong, China}
\author{Andreas Hemmerich}
\affiliation{Institute of Quantum Physics, University of Hamburg, Luruper Chaussee 149, 22761 Hamburg, Germany}
\author{Zhi-Fang Xu}
\email{xuzf@sustech.edu.cn}
\affiliation{Shenzhen Institute for Quantum Science and Engineering and Department of Physics, State key laboratory of quantum functional materials, and Guangdong Basic Research Center of Excellence for Quantum Science, Southern University of Science and Technology, Shenzhen 518055, China}

\begin{abstract}
We report on the experimental observation of classical Brownian motion in momentum space by a Bose-Einstein condensate (BEC) of Rubidium atoms prepared in a hexagonal optical lattice. Upon suddenly increasing the effective atomic mass, the BEC as a whole behaves as a classical rigid body with its center-of-mass receiving random momentum kicks by a Langevin force arising from atom loss and interactions with the surrounding thermal cloud. Physically, this amounts to selective heating of the BEC center-of-mass degree of freedom by a sudden quench, while with regard to the relative coordinates, the BEC is stablized by repulsive atomic interactions, and its internal dynamics is suppressed by forced evaporative cooling induced by atom loss. A phenomenological theory is developed that well explains the experimental data quantitatively.
\end{abstract}

\maketitle

A key insight of Einstein into Brownian motion in $1905$ is that a large classical particle suspended in a thermal environment undergoes rapid momentum randomization~\cite{Einstein1905}. Extending this picture to quantum objects is not only of fundamental interest, but also has many practical applications. As a matter of fact, direct observations of momentum randomization are notoriously difficult~\cite{Raizen2010}, and remain a challenge at low temperatures required for maintaining quantum coherence. Great efforts have been made to investigate the influences of the quantization of Brownian particles~\cite{Leggett1983, Grabert1988, Hanggi2005}. On the other hand, a variety of classical objects, such as solitons~\cite{Kartashov2011} and exciton-polaritons~\cite{Deng2010}, can be formed in quantum many-body systems. This raises the question to what extent Einstein's insight can be carried over to such emergent classical objects. In fact, classical Brownian motion of solitons formed in an ultracold atomic system has been experimentally found to display an anomaly~\cite{Spielman2017}.

Here, we show that a Bose-Einstein condensate (BEC), formed by a large number atoms condensed in the same momentum state, can in fact act as a perfect classical rigid body with regard to its center-of-mass (COM) coordinates. While the internal dynamics of BECs associated with the relative coordinates, such as that of collective excitations, has been studied extensively over decades \cite{Dalfovo1999, Jin1996, Mewes1996, Edwards1996}, little attention has been paid to their COM dynamics. When the system is driven out of equilibrium, say, by a quantum quench, the interaction of the condensate with the surrounding thermal cloud can give rise to intricate condensate COM dynamics, whose properties and consequences have remained largely unexplored.

\begin{figure}
\centering
\includegraphics[width=\linewidth]{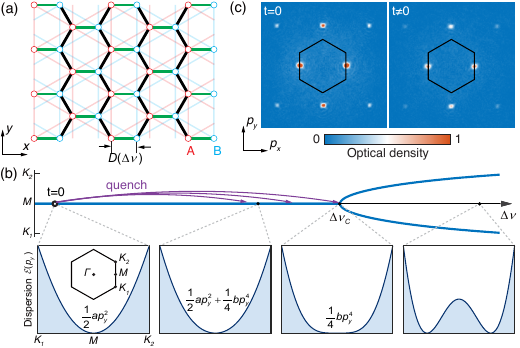}
\caption{(a) A hexagonal optical lattice,  prepared by superposing two triangular sub-lattices A and B, is deformed, by a sudden change of the distance $D(\Delta\nu)$ between adjacent A and B sites in the $x$-direction. (b) The second-band dispersion evolves with $\Delta\nu$. At $t=0$ the system is quenched from some $\Delta\nu$ well below the critical point $\Delta\nu_c$ into the vicinity of (but still below) $\Delta\nu_c$. (c) Both pre- and post-quench momentum distributions exhibit sharp Bragg peaks, indicating the prevalence of strong coherence throughout the quench dynamics. The solid dark honeycombs show the boundary of the first Brillouin zone.}
\label{fig1}
\end{figure}

\begin{figure*}
\centering
\includegraphics[width=\linewidth]{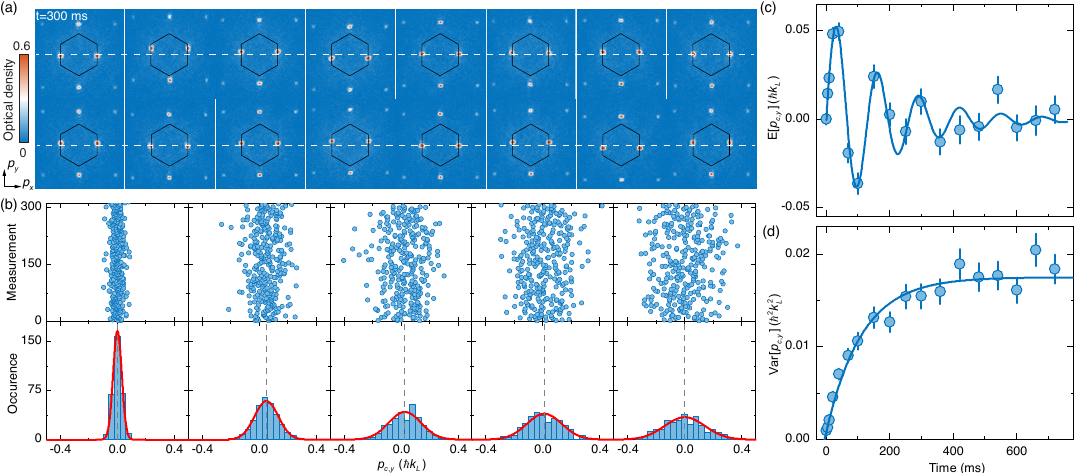}
\caption{(a) The measured momentum distribution of the BEC at $t=300\, \rm ms$ after quench to $\Delta\nu=3.23$ GHz. Each of the 16 panels represents the result from a different experimental run. In all runs two sharp Bragg peaks appear at two equivalent locations at the boundary of the first Brillouin zone, implying condensation at the corresponding momenta. Note that the $y$-component $p_{c,y}$ of the observed momenta display strong fluctuations. Horizontal dashed lines correspond to $p_{y}=0$. The solid black honeycombs indicate the boundary of the first Brillouin zone. (b) The BEC momentum $p_{c,y}$ across different measurements at $t=0,\, 40,\, 150,\, 300$, and $660\, {\rm ms}$, along with the corresponding histogram shown below. The red solid curves indicate Gaussian fits and the vertical dashed lines represent their centers. (c) The time evolution of the average of $p_{c,y}$ over $300$ repeated measurements displays a damped oscillation. (d) The time evolution of the variance of $p_{c,y}$. The error bars represent $68\%$ confidence intervals under Gaussian fitting. Dots denote experimental data and the solid lines denote fitting curves based on Eqs.~(\ref{Mvariance}) and (\ref{eq:mean_measurements}).  In (a-d), the same $\Delta\nu$ is used.}
\label{fig2}
\end{figure*}

We report an experimental observation of classical Brownian motion in the COM momentum of a BEC at an extremely low temperature. The momentum randomization is visualized directly, which is difficult to achieve in conventional experiments on Brownian motion~\cite{Raizen2010}. To be specific, we load a cloud of $^{87}$Rb atoms into the second Bloch band of a deformed hexagonal optical lattice. We find that a sudden quench neither destroys the condensate nor leads to its fragmentation; rather, the condensate as a whole undergoes random momentum scattering so that its COM displays a Brownian motion in momentum space, and at long times the momentum distribution is Maxwell-Boltzmann. A phenomenological theory is developed to describe the observation quantitatively. The observed Brownian motion reveals an unexpected thermalization mechanism in open ultracold atomic systems: Upon application of a quench, on one hand the repulsive interatomic interaction stabilizes the BEC, while on the other hand forced evaporative cooling associated with atom loss suppresses the internal dynamics of the BEC. Consequently, work done during the sudden quench is transferred into heat carried by the COM. This thermalization scenario, finding its origin in random momentum scattering, bears a firm analogy to well-known stochastic heating in plasma physics~\cite{Fermi49, Stix64, Sturrock67, Petrosian12}.

Experimentally, we use an oblate dipole trap and a two-dimensional hexagonal optical lattice~\cite{Wang2021,Liu2022} to confine atoms. The lattice potential is formed by superimposing two triangular sub-lattices, with lattice sites labeled as A and B, respectively. Each sub-lattice is generated by three interfering laser beams, linearly polarized along the $z$-axis and propagating in the $xy$-plane with a wavelength $\simeq 1064\,$nm. We adjust the frequency difference $\Delta\nu$ between two sets of laser beams to control the relative displacement between the two sub-lattices, as illustrated in Fig.~\ref{fig1}(a). This displacement, denoted as $D(\Delta\nu)$, represents the distance between nearest neighbor sites along the $x$-axis. Varying $\Delta\nu$ and the relative depth of the two triangular sub-lattices, we can fine-tune the dispersion of the second Bloch band. In Fig.~\ref{fig1}(b), we illustrate how the second-band dispersion around the $M$ point in the $p_y$ direction transits from a quadratic to a quartic and eventually to a double-well form, as $\Delta\nu$ increases and passes through a critical point $\Delta\nu_c$. The dispersion in the other two directions is always quadratic.

The experiment includes two stages. In the first stage, we load atoms into the second band and implement forced evaporative cooling to induce a BEC around the sole band minimum at the $M$ point, confirmed by time-of-flight (TOF) measurements depicted in Fig.~\ref{fig1}(c), where sharp Bragg peaks indicate pronounced coherence among the entire atomic sample. In the second stage, we suddenly increase $\Delta\nu$ to some value that approaches, but does not exceed $\Delta\nu_c$. Thus, the $M$ point remains as the minimum of the excited band after the quench. Subsequently, the system is allowed to evolve. Throughout this work $t=0$ refers to the beginning of the second stage.

\begin{figure}
\centering
\includegraphics[width=\linewidth]{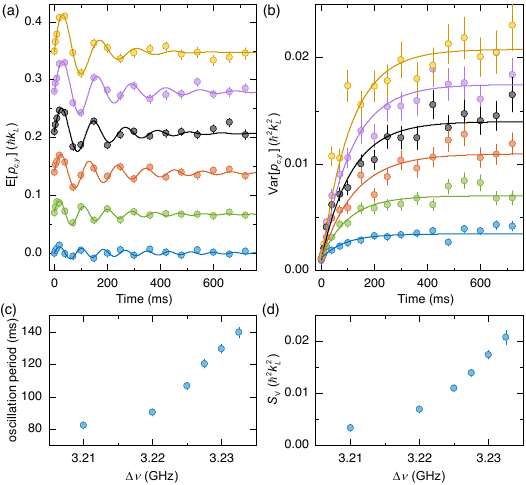}
\caption{(a) Temporal profiles of the mean $p_{c,y}$ for different quench parameters: $\Delta\nu=3.21$, $3.22$, $3.225$, $3.2275$, $3.23$, and $3.2325$ GHz from bottom to top. They are all close to, but below $\Delta\nu_c$. The upper five data sets are shifted upwards to aid legibility. Dots represent experimental data and lines represent the fitting curve of Eq.~(\ref{eq:mean_measurements}). (b) Temporal profiles of the variance of $p_{c,y}$ obtained from the same experimental data sets used in (a), plotted in the same color. Error bars represent $68\%$ confidence intervals for Gaussian fitting.
(c, d) Dependence of the oscillation period and the saturation value of the variance on $\Delta\nu$ obtained by fitting experimental data}.
\label{fig3}
\end{figure}

Upon application of the sudden quench, a certain amount of energy is injected into the system, driving it out of equilibrium. Our measurements indicate that the degree of coherence of the BEC remains high throughout the experiment, as evidenced by sustained sharp Bragg peaks in TOF measurements (cf.~Fig.~\ref{fig1}(c)). The maintenance of coherence is attributed to the repulsive character of the atomic interaction and to the forced evaporative cooling induced by atom loss along the direction of gravity, as observed in previous experiments~\cite{Wang2021, Wang2023}. Moreover, repeating experiments with the same evolution time $t$, we observe significant fluctuations of the momentum distributions. Figure \ref{fig2}(a) shows the momentum distributions derived from TOF measurements at $t=300$ ms after the system is quenched to $\Delta\nu=3.23\, \rm GHz$ close to the critical point. In each single-run, atoms condense around the center momentum $\mathbf{p}_c$, obtained from averaging the single-run momentum distribution around the two equivalent primarily occupied Bragg peaks on the boundary of the first Brillouin zone (solid honeycombs in Fig.~\ref{fig1}(c)), connected by a reciprocal lattice vector.

We perform consecutive experimental runs for more than $300$ iterations for each evolution time, and record ${\bf p}_c \equiv (p_{c,x},p_{c,y})$. The results for the $y$-components, $p_{c,y}$, are plotted in Fig.~\ref{fig2}(b). At time $t=0$, the fluctuations of $p_{c,y}$ are smaller than $\hbar k_L$ by an order of magnitude, suggesting that the initial preparation of the BEC is highly reproducible. As the time increases, the center of the main Bragg peak displays increasing randomness in the momentum space coordinate ($p_{c,y}$). In the extreme case, fluctuations drive $p_{c,y}$ from the $M$ to the $K_1$ (or $K_2$) point at the corner of the first Brillouin zone. The histogram of measurements of $p_{c,y}$ is well fitted by a Gaussian distribution at each instant after the quench. In contrast, we find that the $x$-component, $p_{c,x}$, does not exhibit significant fluctuations. This can also be seen from Fig.~\ref{fig2}(a), where the main Bragg peaks stay at the $K_1$-$K_2$ line throughout the experiment. To analyze fluctuations of $p_{c,y}$, we compute the mean and variance denoted as $\mathrm{E}[p_{c,y}]$ and $\mathrm{Var}[p_{c,y}]$, respectively, with the average taken over repeated measurements for the same evolution time. The results are shown in Fig.~\ref{fig2}(c) and (d). We observe that $\mathrm{E}[p_{c,y}]$ exhibits damped oscillations over time, with an initial amplitude about $5\%$ of $\hbar k_L$. Moreover, $\mathrm{Var}[p_{c,y}]$ increases with time, gradually reaching saturation after several hundred milliseconds of evolution. These observations imply that dissipation is not negligible for our system, and the random motion equilibrates at long time.

Figure~\ref{fig3} shows that momentum fluctuations also occur for other values of $\Delta\nu$. As $\Delta\nu$ approaches $\Delta\nu_c$, the oscillation amplitude and period of $\mathrm{E}[p_{c,y}]$ become larger (Fig.~\ref{fig3}(a) and (c)); simultaneously, the variance of $p_{c,y}$ increases more rapidly and reaches a higher saturation value denoted as $S_{\mathrm{V}}$ (Fig.~\ref{fig3}(b) and (d)). 

\begin{figure}
\centering
\includegraphics[width=\linewidth]{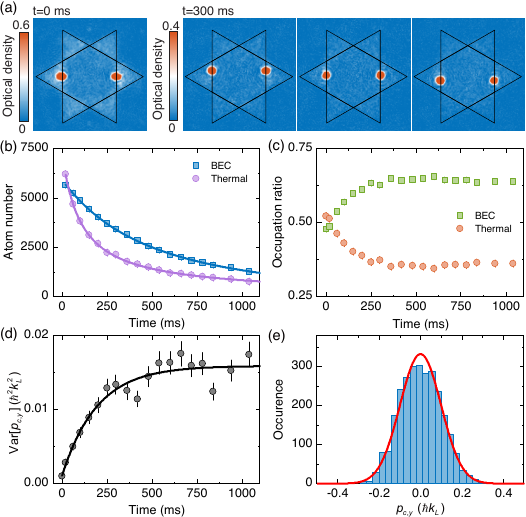}
\caption{(a) Quasi-momentum distributions for $t=0$ and $t=300\, \rm ms$. For the latter time three representative results are shown, which display condensation at different values of $p_{c,y}$.
The inner solid hexagon and the six surrounding triangles denote the first and second Brillouin zones, respectively. (b) Time evolution of the number of condensed (squares) and thermal (circles) atoms in the second Bloch band. The solid lines represent the fitting results based on one- and two-body loss models. (c) Occupation ratios for condensed atoms and thermal atoms in the second band. Error bars in (b) and (c) denote standard error of the mean. (d) Time dependence of the variance of $p_{c,y}$ for condensed atoms. Dots denote experimental data and error bars represent $68\%$ confidence intervals for Gaussian fitting. The solid lines denote fitting curves based on Eq.~(\ref{Mvariance}). (e) Histogram of $p_{c,y}$ at $t=940\,\rm ms$. The red solid curve represents Gaussian fitting. Here $\Delta\nu=3.2275\,\rm GHz$ for all data.
}
\label{fig4}
\end{figure}

In order to distinguish the relative populations and band-specific distributions of the thermal and condensed fractions, we complement in Fig.~\ref{fig4} the TOF data obtained in Figs.~\ref{fig2} and \ref{fig3} by means of a band-mapping technique~\cite{Greiner2001}, that enables the observation of the quasi-momentum distribution. In Fig.~\ref{fig4}(a), we present quasi-momentum spectra at the initial ($t=0$) and late ($t=300\,$ms) evolution time. At $t=0$ atoms mainly occupy the second Bloch band, with a significant fraction of atoms condensed around the $M$ point, evidenced by sharp Bragg peaks. 
Assuming that coherence of thermal atoms can be completely neglected, we approximate their distribution in the second band via band mapping by a uniform distribution, which facilitates separate counts of condensed and thermal atoms~\cite{Nuske2020}. Figure~\ref{fig4}(b) depicts the corresponding temporal evolution of the respective atom numbers. We observe that the loss of thermal atoms is faster than that of the condensed component, resulting in an increased fraction of condensed atoms in the second band, which saturates after several hundred milliseconds (Fig.~\ref{fig4}(c)), indicating thermal equilibrium between thermal atoms and condensed atoms in the second band. Additionally, a notable population of atoms exists in the lowest Bloch band. For our experimental parameters, atoms in the first band primarily occupy $s$-orbitals of deeper lattice sites, spatially segregated from those in the second band, which mainly occupy $s$-orbitals of the shallower lattice sites. Because the contact interaction between first and second band atoms is minimal, thermal atoms in the first band can be ignored. In Fig.~\ref{fig4}(d), we present the time evolution of the variance $\mathrm{Var}\, [p_{c,y}]$, with atoms condensed in the second band being taken into account only, and compare it with the data obtained by TOF spectroscopy \cite{suppl}.
The measurements also show that, in the entire course of time evolution, the momentum distribution remains Gaussian, with the variance increasing with time and saturating at long time. The equilibrium distribution at $t=940\,\rm ms$, obtained from over $3000$ repeated measurements, is depicted in Fig.~\ref{fig4}(e).

The experimental observations suggest that the BEC behaves like a classical object, performing a random motion in momentum space. This is a distinctive feature of Brownian motion. We now turn to the theoretical analysis of this motion and focus on the condensate COM dynamics in the $y$-direction which displays sizable momentum fluctuations, which is described in terms of the phase-space coordinates ($p_{c,y},\,y_c$). Without interactions with incoherent environments, the Hamiltonian of the condensate COM reads $\mathcal{H}=\frac{1}{2}a\,p_{c,y}^{2}+\frac{1}{2}\kappa \, y_c^{2}$. The first term is given by the band dispersion, with $a$ denoting the inverse of the effective atomic mass $m^*$, and the second term accounts for the residual harmonic potential of the dipole trap and the hexagonal optical lattice, with $\kappa$ being nearly constant. In principle there is a quartic $\sim p_{c,y}^{4}$ correction, but since in our experiments the motion does not reach out too far from the band minimum at $M$, its effects are neglected. Effects of the interaction between the BEC and its incoherent environments on the BEC dynamics are two-fold. First, the interaction-induced cooling introduces a frictional force. Second, the observed atom loss from the BEC can introduce a random recoil. In addition, via collisions with the condensate, thermal atoms can also introduce random forces. Taking all these into account, we model the condensate COM dynamics by
\begin{eqnarray}
    \frac{dy_c}{dt}=\frac{p_{c,y}}{m^{*}},\qquad
    \frac{dp_{c,y}}{dt}=-\gamma \, p_{c,y} -\kappa \, y_c+\zeta(t).
    \label{equationofmotion}
\end{eqnarray}
Here $-\gamma \, p_{c,y}$ and $-\kappa \, y_c$ are the frictional force (with $\gamma$ being the friction constant) and the restoring force due to the harmonic potential, respectively. $\zeta(t)$ is a Gaussian white noise that introduces random momentum scattering, whose statistics is completely determined by the autocorrelation: $\mathrm{E}\,[\zeta(t)\zeta(t')]=\Gamma\delta(t-t')$ with $\Gamma$ the noise strength. One can derive the Fokker-Planck equation from Eq.~(\ref{equationofmotion}) and find the full phase-space distribution $f(y_c,p_{c,y};t)$~\cite{Wang1945}. Integrating out $y_c$, we obtain the momentum distribution
\begin{eqnarray}
    f(p_{c,y};t)=\frac{1}{\sqrt{2\pi\mathrm{Var}\,[p_{c,y}]}}\exp\left(-\frac{(p_{c,y}-\mathrm{E}\,[p_{c,y}])^2}{2\mathrm{Var}\,[p_{c,y}]}\right)\,\,
\label{Meq:1}
\end{eqnarray}
governed by the instantaneous mean and variance:
\begin{eqnarray}
   \mathrm{E}\,[p_{c,y}]&=&A_p\, {\rm e}^{-\gamma t/2}\,\sin(\omega_1t+\phi_p), \label{Mmeanvalue}\\
  \mathrm{Var}\,[p_{c,y}]&=&\frac{\Gamma}{2\gamma}\left(1-{\rm e}^{-\gamma t}
  [c_p^2-\frac{c_p\gamma}{2\omega_1}\sin(2\omega_1t+\varphi)]
  \right).\quad
    \label{Mvariance}
\end{eqnarray}
Here, $\omega_1^2=\kappa/m^{*}-\gamma^2/4$, $c^2_{p}=\kappa/(m^*\omega_1^2)$, $\tan\varphi={\gamma/(2\omega_1)}$, and $A_p,\phi_p$ depend on the initial $(p_{c,y},\,y_c)$, whose explicit forms are given in Ref.~\cite{suppl}. Thus $\mathrm{E}\,[p_{c,y}]$ depends on the initial $(p_{c,y},\,y_c)$, but $\mathrm{Var}\,[p_{c,y}]$ does not. 
Below we use Eqs.~(\ref{Meq:1})-(\ref{Mvariance}) to explain the experimental observations quantitatively.

First, $\mathrm{E}\,[p_{c,y}]$ depends on the initial ($p_{c,y},\,y_c$) linearly~\cite{suppl}. Consequently, the average of $\mathrm{E}\,[p_{c,y}]$ over repeated measurements has the same expression as Eq.~(\ref{Mmeanvalue}) (thus we use the same symbol), except that in $A_p,\,\phi_p$ the initial $(p_{c,y},\,y_c)$ are replaced by their average over measurements, denoted as $(\overline{p_0},\overline{y_0})$~\cite{suppl}:
\begin{eqnarray}
{\rm E}\,[p_{c,y}]
=A_p(\overline{y_0},\overline{p_0})\,{\rm e}^{-\frac{\gamma t}{2}}\sin[\omega_1 t+\phi_p(\overline{y_0},\overline{p_0})].
\label{eq:mean_measurements}
 \end{eqnarray}
Eq.~(\ref{eq:mean_measurements}) well fits measurements (Figs.~\ref{fig2}(c) and \ref{fig3}(a)),
and shows that the damped oscillation has angular frequency $\omega_1$ and decay rate $\gamma/2$. In addition, quenching closer to the critical point leads to larger mass $m^*$, and using the expression of $\omega_1$ we find that this results in the enhancement of the oscillation period, consistent with measurements
in Fig.~\ref{fig3}(a)~\cite{suppl}. Furthermore, in our experiments the sinusoidal oscillation in Eq.~\eqref{Mvariance} is weak;
moreover, to prove or disprove such a weak oscillation much more data are required which, unfortunately, is beyond our experimental reach. Thus we simplify Eq.~(\ref{Mvariance}) to $\mathrm{Var}\,[p_{c,y}]=\frac{\Gamma}{2\gamma}(1-c_p^2\,{\rm e}^{-\gamma t})$, which well fits 
measurements (Figs.~\ref{fig2}(d) and \ref{fig3}(b)).

Figures \ref{fig2}(b) and \ref{fig4}(e) further show that the full momentum statistics, namely, the Gaussian distribution given by Eq.~(\ref{Meq:1}) agrees well with the experimental data throughout the evolution. By Eq.~(\ref{Mvariance}), at the time scale $t_{\rm Th}=1/(2\gamma)$ a crossover from $\mathrm{Var}\,[p_{c,y}]=\Gamma t$ to saturation: $\mathrm{Var}\,[p_{c,y}]=\Gamma/(2\gamma)$ occurs. The linear growth suggests that for $t\ll t_{\rm Th}$ the condensate COM undergoes a diffusive motion in the momentum space. Since a BEC atom has mean kinetic energy $\mathrm{Var}\,[p_{c,y}]/(2m^*)$, it gets heated up with a constant rate $\Gamma/(2m^*)$ by random momentum scattering. For longer times, the cooling comes into play. According to Eq.~(\ref{equationofmotion}) the cooling rate is  $\gamma\mathrm{Var}\,[p_{c,y}]/m^*$, which balances the heating rate only if $\mathrm{Var}\,[p_{c,y}]$ attains the saturation value $\Gamma/(2\gamma)$: This implies that the COM of the BEC reaches thermal equilibrium at $t\sim t_{\rm Th}$ with Eq.~(\ref{Meq:1}) being a Maxwell-Boltzmann distribution, as if the BEC as a whole was a classical object. The kinematic temperature is
\begin{equation}
T=N_{\rm BEC}\frac{ \mathrm{Var}[p_{c,y}(t\gtrsim t_{\rm Th})]}{k_Bm^*}=\frac{N_{\rm BEC}\Gamma}{2k_B\gamma m^*},
\label{energyequipartition}
\end{equation}
with $k_B$ being the Boltzmann constant. Accordingly, for fixed $N_{\rm BEC}$ (condensed atom number) and $T$, the enhancement of $m^*$ leads to the enhancement of the saturation value $\Gamma/(2\gamma)$, consistent with the measured temporal profiles of $\mathrm{Var}[p_{c,y}]$ in Fig.~\ref{fig3}(b). As detailed in Ref.~\cite{suppl}, we can determine ${\Gamma/(2\gamma)}$ from experimental data and $m^*$ from calculations. Substituting them and the estimation $N_{\rm BEC}\simeq 10^3$ into Eq.~(\ref{energyequipartition}), we find $T$ varies in $35$-$60$ nK for $\Delta\nu$ considered in Fig.~\ref{fig3}. On the other hand, we expect that $T$ makes no reference to the direction. In the $z$-direction, it is the order of the trap depth along the gravity direction~\cite{Wang2021}, which is about $40$ nK. We see the two estimations for $T$ are consistent.

Our thermalization scenario differs from other heating processes in quantum gases. 
Notably, owing to the strong interplay between dissipation and quantum coherence, the BEC is not destroyed upon quench. Opposite to this, in previous experiments on quenching closed (thus non-dissipative) systems~\cite{Parker2013,Clark2016,Nicklas2015}, the BEC is destroyed and structures such as domain walls are formed. Additionally, at ultralow temperatures, it is conceivable that matter wave effects may result in nontrivial heating processes. Indeed, a profile similar to Fig.~\ref{fig2}(d) was experimentally observed in driven cold quantum gases \cite{Raizen95}, but that is due to Anderson localization \cite{Casati79,Fishman82,Tian04,Tian10} and the stationary distribution is exponential.

Summarizing, we have shown that during quench dynamics a BEC interacting with a surrounding thermal cloud and subjected to evaporative cooling induced by atom loss can display classical Brownian motion in momentum space. The ensuing thermalization of the condensate COM degree of freedom differs from various scenarios for thermalization of quantum systems \cite{Rigol16, Borgonovi16, Nandkishore2015, Ueda2020}. 
In view of the discovery of quantum turbulence in cold atomic gases \cite{Henn2009,Hadzibabic16}, the origin of stochasticity can be more complex than Gaussian noise assumed in this work, which may potentially lead to rich new non-equilibrium phenomena in cold atomic systems.

We thank Hui Zhai, Congjun Wu, Shuai Yin, Qi Zhou, Yu Chen, and Masahito Ueda for useful discussions. This work was supported by the National Key R\&D Program of China (Grant No.~2022YFA1404103), NSFC (Grants No.~12274196, No.~12304289, and No.~92476101), and funds from Guangdong province (Grants No.~2019QN01X087 and No.~2019ZT08X324); C.T. acknowledges support from NSFC (Grants No.~11925507, No.~12475043 and No.~12047503); S.Z. acknowledges support from HK GRF (Grant No.~17306024), CRF (Grants No.~C6009-20G and No.~C7012-21G), and a RGC Fellowship Award No. HKU RFS2223-7S03; A.H. acknowledges support from the Cluster of Excellence CUI: Advanced Imaging of Matter of the Deutsche Forschungsgemeinschaft (DFG) - EXC 2056 - project ID 390715994.

\renewcommand\thefigure{S\arabic{figure}}
\setcounter{figure}{0}
\renewcommand\theequation{S\arabic{equation}}
\setcounter{equation}{0}
\makeatletter
\newcommand{\rmnum}[1]{\romannumeral #1}
\newcommand{\Rmnum}[1]{\expandafter\@slowromancap\romannumeral #1@}

\newpage
\onecolumngrid

{
	\center \bf \large
	Supplemental Materials
	\vspace*{0.1cm}\\
	\vspace*{0.0cm}
}

\section*{Experimental system}
Our experiment starts with a $^{87}$Rb Bose-Einstein condensate consisting of typically $6.8\times10^4$ atoms in the $|F=1, m_F=-1\rangle$ state. The condensate is prepared in an oblate optical dipole trap with trapping frequencies $\lbrace \omega_x,\omega_y,\omega_z \rbrace=2 \pi \times \lbrace 26.4, 26.5, 69.5\rbrace\,$Hz. Atoms are subsequently loaded into the deformed hexagonal lattice, which is realized by superimposing two sets of triangular optical lattices with lattice sites denoted as A and B respectively. They are created by three running-wave laser beams linearly polarized parallel to the $z$-axis and intersecting at an angle of $120^{\circ}$ in the $xy$-plane. Each laser beam comprises two frequency components of $\nu_1$ and $\nu_2$ with their difference defined as $\Delta\nu=\nu_1-\nu_2$. Both wavelengths are $\lambda \simeq 1064$ nm. The lattice potential is thus given by
\begin{eqnarray}
V(\mathbf{r})&=&-V_A\left[3+2\sum\nolimits_{j}\cos \left( \mathbf{b}_j\cdot \mathbf{r}\right)  \right]  \nonumber\\
&&-V_B\left[3+2\sum\nolimits_{j}\cos \left( \mathbf{b}_j\cdot \mathbf{r} -\Delta\eta_j\right)  \right].
\label{bn}
\end{eqnarray}
Here, the vectors $\mathbf{b}_j$ are defined as $\mathbf{b}_1=\mathbf{k}_1-\mathbf{k}_2$, $\mathbf{b}_2=\mathbf{k}_2-\mathbf{k}_3$, and $\mathbf{b}_3=\mathbf{k}_3-\mathbf{k}_1$ with the wave vectors ${\bf k}_1=k_L(-\sqrt{3}/2,1/2)$, ${\bf k}_2=k_L(\sqrt{3}/2,1/2)$, and ${\bf k}_3=k_L(0,-1)$, where $k_L=2\pi/\lambda$ and $j=1,2,3$. $\Delta\eta_{j}=2\pi\Delta\nu \Delta L_j/c$ determine the difference of the centers of the two triangular lattices. Experimentally, each set of triangular lattice, A or B, is created by three interfering laser beams. The optical path differences are defined as $\Delta L_1=L_1-L_2$, $\Delta L_2=L_2-L_3$ and $\Delta L_3=L_3-L_1$, where $L_j (j=1,2,3)$ represents the optical path of the $j$-th laser beam from the splitting point to the center of the lattice. For more detailed information, we refer to Refs.~[17,18]. We choose $(\Delta L_1,\Delta L_2,\Delta L_3)\simeq(-6.12,3.06,3.06) \, \rm cm$ by changing the optical paths of the laser beams forming the triangular lattices. Since $\Delta L_2=\Delta L_3$, tuning $\Delta\nu$ thus changes the separation of the centers of the two triangular lattices along the $x$-axis. In the case of $\Delta\nu=3.263\,$GHz, $V(\mathbf{r})$ becomes a regular hexagonal lattice. For other cases, a deformed hexagonal lattice is realized.

\section*{State preparation and detection}
Initially, atoms are loaded into the lowest $s$-orbital of the deformed hexagonal optical lattice with $\Delta\nu=3.19\,$GHz and $(V_A, V_B)=(7.99, 5.16)\, E_{\rm{R}}$, where $E_\mathrm{R}=h^2/2m\lambda^2$ is the recoil energy. By quickly changing $(V_A, V_B)$ to $(7.32, 7.93)\,E_{\rm{R}}$ in 0.1 ms, the atoms are transferred to the second Bloch band of the lattice and, via active cooling, are later recondensed in the minimum of the second band, located at the $M$ point. In order to ensure efficient and continuous cooling of the atoms, the intensity of the optical dipole trap is linearly reduced to a certain value within 15 ms, and the final value is carefully optimized to provide an scattering channel for the thermal atoms to escape from the trap along the direction of gravity via collisions. After a holding time of 105 ms, we suddenly change $\Delta\nu$ from 3.19 GHz to a value close to, but not beyond the critical point of the effectively ferromagnetic quantum phase transition. This operation is completed within 0.5 ms. In the following dynamics, we observe the center-of-mass oscillation and fluctuations of the condensate in momentum space over time, centered around the energy minimum of the second Bloch band, using time-of-flight spectroscopy or a band mapping technique.

\section*{Band dispersion engineering}
\begin{figure*}
	\centering
	\includegraphics[width=\linewidth]{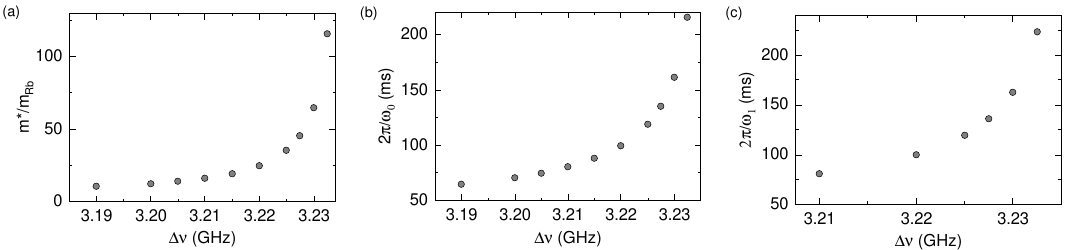}
	\caption{(a) The effective mass $m^*$ for the second Bloch band along the $y$-direction is plotted versus the frequency difference $\Delta\nu$ with $(V_A, V_B)=(7.32, 7.93)\,E_{\rm{R}}$. (b) The estimated oscillation period $2\pi/\omega_0$ with the trapping frequency approximately chosen as $2\pi\times 50\, \rm Hz$. (c) The oscillation period $2\pi/\omega_1$. Here $\omega_1^2=\omega_0^2-\gamma^2/4$. The corresponding values of $\gamma$ are determined from experimental data fitting.}
	\label{sfig1}
\end{figure*}

In our experiment, we consider a situation with different final lattice depths among the A and B sites with $(V_A, V_B)=(7.32, 7.93)\,E_{\rm{R}}$. When the atoms are transferred into the second Bloch band, they mainly populate the shallow $s$-orbitals of the A sites. This leads to an extremely long lifetime. Via tuning $\Delta\nu$, we can significantly change the band dispersion along the $y$ direction, which is shown in Fig.~1(b). In this case, the critical point is estimated to be located around $\Delta\nu_c=3.235\, \rm GHz$. To characterize the band dispersion along the $y$ direction, we fit it with $\mathcal{E}=\mathcal{E}_0+\frac{1}{2}a\,p_y^2+\frac{1}{4}b\,p_y^4$ along the line connecting the $K_1$ and $K_2$ points. When $\Delta\nu<\Delta\nu_c$, $a>0$, while in the opposite case, the band structure shows a double-well dispersion with $a<0$. Focusing on the case with $\Delta\nu<\Delta\nu_c$, we numerically calculate the corresponding parameter of $a$ to evaluate the effective mass along the $y$ direction denoted as $m^*=1/a$. Details are shown in Fig.~\ref{sfig1}(a). While the overall effective trapping frequency due to the lattice potential is difficult to measure directly, we numerically calculate its value by using the measured lattice depth and other relevant laser parameters such as beam waists and intensities. As a result, when the contributions from both the dipole trap and the optical lattice are taken into account, we can determine the overall trapping potential along the $y$-direction approximately, yielding an estimated frequency of $2\pi \times 50\,\rm Hz$.
This enables us to roughly estimate the oscillation period $2\pi/\omega_0$, where $\omega_0\equiv\sqrt{\kappa/m^*}=2\pi\times 50\sqrt{m_{\rm Rb}/m^*}\,\rm Hz$ and $m_{\rm Rb}$ is the atom mass of $^{87}$Rb. Details are shown in Fig.~\ref{sfig1}(b). Furthermore, using the value of $\gamma$ from experimental data fitting, we caculate $\omega_1$ shown in Fig.~\ref{sfig1}(c). The obtained oscillation period $2\pi/\omega_1$ is consistent with experimentally observed oscillation period for the mean value of momentum along the $y$ direction shown in Fig.~3(c).

Note that what is essential to our work is the continuous change in the effective mass $m^*$, and the presence of a critical point, at which the band dispersion in the $y$-direction changes from quadratic to quartic. In our experiment, it is convenient to implement such modification of the band dispersion by tuning the frequency difference. For other experimental setups, the same band dispersion modification might be realized by tuning other parameters. When the atom loss-induced cooling is further under controls, similar dynamics is expected to occur.

\section*{Comparisons of COM temperature by different methods}

\begin{figure}
	\centering
	\includegraphics[width=5.7cm]{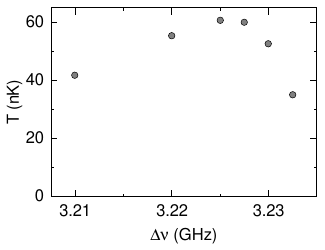}
	\caption{The kinematic temperature for distinct frequency differences $\Delta\nu$ is estimated by using Eq.~(6) in the main text.}
	\label{sfig2}
\end{figure}

On one hand, by using Eq.~(6) in the main text we can estimate the kinematic temperature $T$ of the condensate COM. For the frequency difference $\Delta\nu$ ranging from $3.21$ GHz to $3.2325$ GHz, we find that $T$ varies from $35$ nK to $60$ nK (Fig.~\ref{sfig2}). On the other hand, it is reasonable to expect that the kinetic temperature is independent of the direction. When the thermal equilibrium in the $z$-direction is established, the kinetic temperature is the order of the trap depth along the gravity direction, which is about $40$ nK based on our experimental parameters. This is consistent with that given by Eq.~(6) in the main text.

\section*{Momentum fluctuations in $\boldsymbol{x}$-direction}
\begin{figure}[h]
\centering
\includegraphics[width=11.0cm]{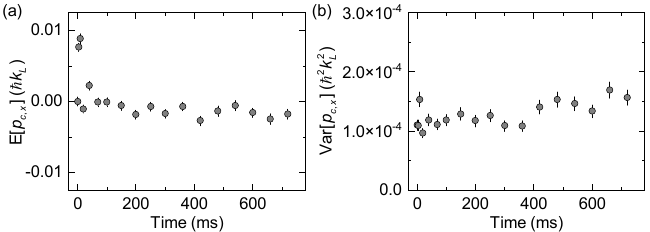}
\caption{The measurement of the time evolution of the mean (a) and variance of the BEC COM momentum in the $x$ direction. The experimental protocol and parameters are the same as those in Fig.~2.}
\label{sfig3}
\end{figure}

In Fig.~\ref{sfig3} we show the experimental results for the fluctuations of the $x$-component of the COM momentum of BEC, denoted as $p_{c,x}$.
Fig.~\ref{sfig3}(a) is the temporal profile of the mean and Fig.~\ref{sfig3}(b) the variance. The experimental protocol and parameters are exactly the same as those in Fig.~2. We see clearly that fluctuations in the $x$-direction are smaller than those in the $y$-direction by over two orders of magnitude.

\section*{Time evolution of atom number in different bands}

\begin{figure}[h]
\centering
\includegraphics[width=7cm]{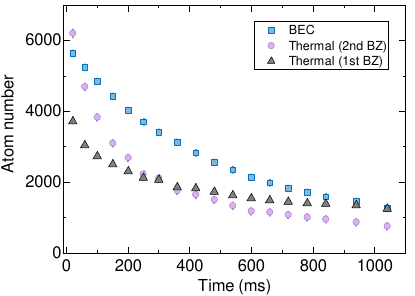}
\caption{The time evolution of the number of thermal atoms in the first and second BZ, as well as the atom number of BEC in the second BZ. $\Delta\nu=3.2275$ GHz.}
\label{sfig4}
\end{figure}

Band mapping analysis allows us to detect the atom number in the first and the second Brillouin zone (BZ). In Fig.~\ref{sfig4} we plot the experimental results for the time evolution of the number of thermal atoms in the first and second BZ, and also atom number of BEC in the second BZ.

\section*{Determination of error bars}
In our data analysis of measured $p_{c,y}$, at each time we fit over $300$ measured data points to a Gaussian distribution of the form: 
$f(p_{c,y})={1\over \sqrt{2\pi \sigma^2}}  e^{-(p_{c,y}-\mu)^2/(2\sigma^2)}$ using the fitdist function of MATLAB, which employs maximum likelihood estimation. Then $\mu$ and $\sigma^2$ give the mean and the variance of $p_{c,y}$, respectively. 
The error bars of them represent $68\%$ confidence intervals obtained from Gaussian fitting. They were computed via MATLAB's paramci function, based on Student's t-test and Chi-squared test.

\section*{Comparison between TOF and band-mapping data}
\begin{figure}[h]
\centering
\includegraphics[width=7cm]{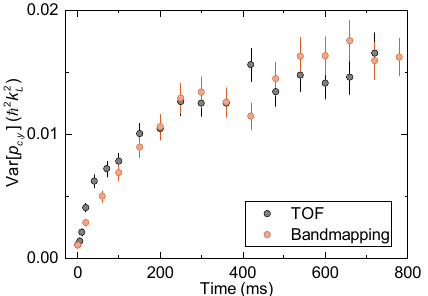}
\caption{The time evolution of the variance of $p_{c,y}$ obtained from TOF and band-mapping measurements at $\Delta\nu=3.2275$ GHz.}
\label{sfig5}
\end{figure}

To compare the data obtained from the TOF and band-mapping techniques, in Fig.~\ref{sfig5} we have plotted Fig. 4(d) alongside the data (black dots) from Fig. 3(b), both of which correspond to
$\Delta\nu=3.2275\,\rm GHz$. The two data sets are in good agreement, which supports the conclusion that the two measurement methods yield consistent results.

\section*{Random force induced by atom loss}
There are different sources giving rise to the random force. Here we point out a mechanism, which is highly relevant to our experiments but yet remains largely unexplored in the literature. That is, atoms can escape from the condensate, and introduce a random recoil. As observed in experiments, the atoms mostly condense at a single quasi-momentum. So a condensed atom carries in $y$-direction a momentum $p_{c,y}$ on the average. Suppose that at time $t$ the BEC consists of $N(t)$ atoms. Then it carries an instantaneous total momentum: $N(t)\,p_{c,y}(t)$. Let an small amount of atoms, $N_L(t)\equiv N(t)-N(t+\delta t)$, be lost from BEC during a small time interval $\delta t$, each of which carries a momentum $p_L(t)\equiv p_{c,y}(t)+\xi(t)$. Here $\xi(t)$ is random and introduces a random recoil to the condensate. Since the total momentum is conserved, we have
\begin{eqnarray}
\label{eq:1}
N(t+\delta t)p_{c,y}(t+\delta t)+N_L(t)p_L(t)=N(t)\left(p_{c,y}(t)-\kappa y_c\delta t\right),
\end{eqnarray}
where $y_c$ denotes the mean position of a condensed atom in the $y$-direction, and $-N(t)\kappa y_c\delta t$ is the total momentum input by the external harmonic trap. Setting $\delta t\rightarrow 0$ we obtain
\begin{eqnarray}
\frac{dp_{c,y}}{dt}=-\kappa y_c+\frac{d\ln N}{dt}\,\xi.
\label{eq:2}
\end{eqnarray}
This equation is difficult to solve because the random recoil $\xi(t)$ is coupled to the atom loss, i.e. $\frac{d\ln N}{dt}$. To overcome this difficulty we instead  treat the second term as a whole as a random force $\zeta(t)\equiv \frac{d\ln N}{dt}\,\xi(t)$. So Eq.~(\ref{eq:2}) is simplified to
\begin{eqnarray}
\frac{dp_{c,y}}{dt}=-\kappa y_c+\zeta (t).
\label{eq:3}
\end{eqnarray}
With the friction force $\gamma p_{c,y}$ being included, we then obtain Eq.~(1) of the main text. As said above, there are other sources of random force such as thermal atoms. They add together, however, and thus do not change the form of this equation of motion.

\section*{Friction force due to cooling}
In our system there is a cooling mechanism, which enables transferring kinetic energy from the $xy$ plane to the $z$-axis via collisions. This opens an efficient channel for evaporation of atoms with high kinetic energy stored in the $z$-direction aided by gravity. This active cooling may serve as an origin for the friction force applied to the BEC, accounted by the first term in Eq.~(1) of the main text.

\section*{Momentum distribution}
For simplicity we assume that $\zeta$ is a Gaussian white noise with $\langle\zeta(t)\rangle=0$ and $\langle\zeta(t)\zeta(t')\rangle=\Gamma\delta(t-t')$. We can derive a Fokker-Planck equation for the Langevin model described by Eq.~(1) in the main text, from which we find the instantaneous phase-space distribution function in the phase space coordinates $(y_c, p_{c,y})$,  read
\begin{equation}\label{eq:4}
  f(y_c,p_{c,y};t)=e^{t{\cal L}_{\rm FP}}\delta(y_c-y_0)\delta(p_{c,y}-p_0),
\end{equation}
where the operator
\begin{equation}\label{eq:5}
  {\cal L}_{\rm FP}=-\frac{p_{c,y}}{m^*}\partial_{y_c} +\partial_{p_{c,y}}\left(\gamma p_{c,y}+\omega_0^2 y_c\right)+{\Gamma\over 2}\partial_{p_{c,y}}^2,
\end{equation}
and $(p_0,\,y_0)$ are the initial coordinates.
Define a two-component complex vector $\boldsymbol{z}=(z_1,z_2)^{\rm T}$, whose components are $z_{1,2}=p_{c,y}/m^*+a_{1,2}y_c$, with $a_{1,2}=\gamma/2\pm i\omega_1$. Then we can write Eq.~(\ref{eq:4}) explicitly, which is
\begin{eqnarray}
\label{eq:6}
f(y_c,p_{c,y};t)={-2i\omega_1}\int {d\boldsymbol{\eta}\over (2\pi)^2}{\rm e}^{i\boldsymbol{\eta}^{\rm T}(\boldsymbol{z}-\boldsymbol{z}_{0}(t))-{1\over 2}\boldsymbol{\eta}^{\rm T}A(t)\boldsymbol{\eta}}.\qquad
\end{eqnarray}
Here
\begin{eqnarray}
\label{eq:7}
\boldsymbol{z}_{0}(t)=\left(
\begin{array}{c}
(p_0/m^*+a_1 y_0){\rm e}^{-a_2t}\\
(p_0/m^*+a_2 y_0){\rm e}^{-a_1t}\\
\end{array}
\right),
\end{eqnarray}
and
\begin{equation}\label{eq:8}
  A(t)={\Gamma\over \gamma m^{*2}}\left(
                                    \begin{array}{cc}
                                      {1-{\rm e}^{-(1-is_p)\gamma t}\over 1-is_p} & 1-{\rm e}^{-\gamma t} \\
                                      1-{\rm e}^{-\gamma t} & {1-{\rm e}^{-(1+is_p)\gamma t}\over 1+is_p} \\
                                    \end{array}
                                  \right),
\end{equation}
with $s_p=2\omega_1/\gamma$, and $\boldsymbol{\eta}=(\eta_1,\eta_2)^{\rm T}$.

The expression (\ref{eq:6}) allows us to perform the integral over $y_c$ and find the momentum distribution $f(p_{c,y};t)$. The result is
\begin{eqnarray}
\label{eq:9}
f(p_{c,y};t)=\int {d\eta\over 2\pi}{\rm e}^{i\eta[p_{c,y}-
F_1(y_0,p_0;t)]}
{\rm e}^{-
F_2(t)\eta^2},\,\,\,\,\,\,
\end{eqnarray}
where
\begin{eqnarray}
F_1(y_0,p_0;t)&=&{{\rm e}^{-\frac{\gamma t}{2}}\over \omega_1}\left[({1\over 2}\,p_0\gamma+\kappa y_0)\sin(\omega_1 t)-p_0\omega_1\cos(\omega_1 t)\right],\label{eq:10}\\
F_2(t)&=&\frac{\Gamma}{4\gamma}\left(1-{\rm e}^{-\gamma t}
  [c_p^2-\frac{c_p\gamma}{2\omega_1}\sin(2\omega_1t+\varphi)]
  \right),\label{eq:13}
\end{eqnarray}
where $\omega_1^2=\omega_0^2-\gamma^2/4$, $c_{p}=\omega_0/\omega_1$, and $\tan(\varphi)=1/s_p$. Integrating out $\eta$, we obtain
\begin{eqnarray}
    f(p_{c,y};t)=\frac{1}{\sqrt{4\pi F_2(t)}}\exp\left(-\frac{(p_{c,y}-F_1(y_0,p_0;t))^2}{4F_2(t)}\right).
\label{eq:11}
\end{eqnarray}
Let us introduce two functions defined as
\begin{equation}\label{eq:17}
  A_p(y,p)\equiv \sqrt{({1\over 2}\,p\gamma+\kappa y)^2/\omega_1^2+p^2},\quad \phi_p (y,p)=-\arctan \frac{p\omega_1}{{1\over 2}\,p\gamma+\kappa y}.
\end{equation}
Then we can rewrite $F$ as
\begin{equation}\label{eq:12}
F_1(y_0,p_0;t)=A_p(y_0,p_0)\,{\rm e}^{-\frac{\gamma t}{2}}\sin[\omega_1 t+\phi_p(y_0,p_0)].
\end{equation}
This implies that $A_p(y_0,p_0)$ and $\phi_p(y_0,p_0)$ give respectively the amplitude and the initial phase of a damped (monochromatic) oscillation. For the distribution (\ref{eq:11}), the first and second moments are
\begin{eqnarray}
{\rm E}\,[p_{c,y}] &=& \int dp_{c,y}\, f(p_{c,y};t)\,p_{c,y}=F_1(y_0,p_0;t),\label{eq:14}\\
{\rm Var}\,[p_{c,y}] &=& \int dp_{c,y}\, f(p_{c,y};t)\,(p_{c,y}-{\rm E}\,[p_{c,y}])^2=2F_2(t),\label{eq:15}
\end{eqnarray}
from which Eqs.~(2)-(4) in the main text follow immediately.

Note that $F_1$ depends on $p_0,\,y_0$, while $F_2$ is independent of $p_0,\,y_0$. Therefore, in each measurement, since the initial
phase coordinates $p_0,\,y_0$ are fixed, the time evolution of the Gaussian distribution (\ref{eq:11}) is completely fixed. At measurement time $t$, the mean, i.e. the center of distribution $F_1$, depends on $p_0,\,y_0$, whereas the variance, i.e. $F_2$, does not. Repeating the measurements, we obtain an ensemble of temporal profiles ${\rm E}\,[p_{c,y}]$, which (at given time) fluctuates from measurement to measurement. Let us perform the ensemble average (denoted as $\overline{\cdots}$). Since by Eq.~(\ref{eq:10}) the dependence of $F_1$ on $p_0,\,y_0$ is linear, and all parameters $\gamma,\,\kappa,\,\omega_1$ are independent of $p_0,\,y_0$, we have
\begin{eqnarray}
\overline{F_1(y_0,p_0;t)}
&=&{{\rm e}^{-\frac{\gamma t}{2}}\over \omega_1}\left[({1\over 2}\,\overline{p_0}\gamma+\kappa \overline{y_0})\sin(\omega_1 t)-\overline{p_0}\omega_1\cos(\omega_1 t)\right],\nonumber\\
  &=& A_p(\overline{y_0},\overline{p_0})\,{\rm e}^{-\frac{\gamma t}{2}}\sin[\omega_1 t+\phi_p(\overline{y_0},\overline{p_0})].\label{eq:16}
 \end{eqnarray}
This expression differs from Eq.~(\ref{eq:14}) only in that in $A_p$ and $\phi_p$ the arguments $(y_0,p_0)$ are replaced by $(\overline{y_0},\overline{p_0})$.
Therefore, we use the same symbol ${\rm E}\,[p_{c,y}]$ for the temperal profiles of ${\rm E}\,[p_{c,y}]$ with and without averaged over reapeated measurements.
 Eq.~(\ref{eq:16}) gives the theoretical curve in Fig.~2(c) and Fig.~3(a) of the main text.

\end{document}